\documentclass[epj]{svjour}
\usepackage{graphics}
\usepackage{amsmath}
\usepackage{amsfonts}
\usepackage{amssymb}
\usepackage{epsfig}
\usepackage{verbatim}
\newcommand{\feyndag}[1]{\ensuremath{\not{\hspace{-0.75mm} #1}}}
\begin{document}
\def\Id{{\rm 1\kern-.3em I}}
\title{Generalized parton distributions of pseudoscalar mesons in a
  covariant constituent quark model} 
\author{A. Van Dyck\inst{1} \and T. Van
 Cauteren\inst{1}\thanks{\email{Tim.VanCauteren@UGent.be}}
 \and J. Ryckebusch\inst{1} \and B. C. Metsch\inst{2}
}                     
\institute{Department of Subatomic and Radiation Physics, Ghent
  University, Proeftuinstraat 86, B-9000 Gent, Belgium \and
  Helmholz-Institut f\"ur Strahlen- und Kernphysik, Nu\ss{}allee
  14-16, D-53115 Bonn, Germany}   
\date{Received: date / Revised version: date}
%
\abstract{
  The isoscalar twist-two generalized parton distributions (GPDs) of
  the pion and the kaon are calculated in a Poincar\'e covariant
  Bethe-Salpeter constituent quark model. Results are presented for
  several values of the parameters $\xi$ and $t$. The results satisfy
  the form factor constraints and the polynomiality condition. For the
  pion GPD, also the isospin symmetry constraint is fulfilled. The
  influence of kinematical variables and model parameters on the
  support of the GPDs is investigated. To this end, the strength
  parameters and quark masses of the constituent quark model are
  artificially varied.
} 
\maketitle
%
%
\section{\label{sec:intro}Introduction}
Unraveling the substructure of hadrons is a challenging issue.
Whereas the theory of QCD provides us
with the equations governing the quark-gluon dynamics, the exact way
in which they form colorless bound states in the non-perturbative
regime remains elusive to date.
It is expected that the endeavor of measuring and computing
generalized parton distribution functions (GPDs) will contribute
significantly to a full description of hadron
structure~\cite{Boffi:2007yc}. GPDs are a natural unification of form
factors (FFs) and parton distribution functions (PDFs) within one
framework. They describe the non-perturbative part of the deeply
virtual Compton scattering (DVCS) and hard exclusive meson production
(HEMP) processes.

A calculation of GPDs from first QCD principles is beyond reach
at this moment. Therefore, numerous model calculations have been presented in
the last
years~\cite{noguera:bsapp,luik:gpd,tibmil1,tibmil2,tibmil3,kiss,choi1,choi2,vandyckvancauteren:proceedings}. In
this work, we present results for the pion and kaon GPDs computed
within the framework of the Bethe-Salpeter model developed in
refs.~\cite{koll00,resag95,muenz94,muenzthesis}. The pseudoscalar
ground-state mesons have a simple valence-quark substructure and
possess only one helicity-conserving GPD, which makes the calculation
less cumbersome than for nucleons. Yet, such a calculation
provides insight in the dependence of the GPDs on hadron structure in
the non-perturbative regime.

In order to compare the GPDs computed in a phenomenological model with
the data from deeply inelastic scattering experiments, a
$Q^2$-evolution needs to be performed \cite{jaffe:q2}. The evolution
equations depend highly on the kinematic region of the process. More
specifically, one can distinguish between the DGLAP and ERBL regions,
where the $Q^2$-evolution is governed by the DGLAP and the ERBL
equations, respectively. Parton model constraints make GPDs vanish
outside these two regions \cite{diehl:review,jaffe:pdftwist4}. A model
which resolves the ERBL and the DGLAP regions and has a vanishing GPD
otherwise, is said to have the correct support.

In ref.~\cite{vandyckvancauteren:letter} we have demonstrated that a
support problem may arise in dynamic quark models based on the BS
approach if the interaction depends on the minus component of the
relative momentum between the constituent quark and antiquark. Note
that this condition is necessary but not sufficient and that a support
problem does not necessarily arise when the interaction depends on the
minus component of the relative momentum. Furthermore, we provided
representative results computed within the explicitly Poin\-car\'e
covariant constituent quark mo\-del developed by the Bonn group
\cite{koll00}, where a support problem did arise.

In this work, we focus on the influence of the meson binding energy
on the support. The pion and the kaon are ideal for such a study because
the former is a very deeply and the latter a moderately deeply bound
state of quark and antiquark. Both the absolute binding energy and meson
mass are altered by changing the input quark masses and effective
interaction strength. Yet the degree of support violation depends
mostly on the relative binding energy. Results will be shown for the
isoscalar twist-two quark GPD of the pion and the kaon at different
relative binding energies and for different values of the squared
fourmomentum transfer $t$ and relative plus-momentum transfer $\xi$. 

This paper is organized as follows. Section \ref{sec:kine} presents
the definition of the GPD and the relevant kinematics. The Bonn model
is introduced in sect. \ref{sec:bonnmodel}, while
sect. \ref{sec:gpd_in_bm} is devoted to the calculation of GPDs in
this model. In this section, we will also introduce the three model
variants which will be used in sect. \ref{sec:results&disc}. There, we
show our results for the pion GPD. A summary and an outlook to future
research are given in sect. \ref{sec:Sum&concl}.

\section{\label{sec:kine}Definition and Conventions}
\begin{figure}
\centering
\includegraphics[width=0.8\columnwidth]{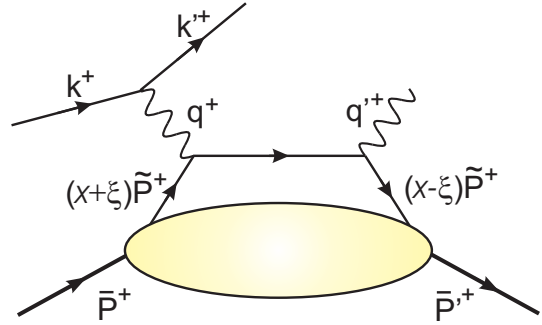}
\caption{The Feynman diagram of DVCS on the light cone, with the
  definition of the kinematical variables $x$ and $\xi$.}
\label{Fig:kinematics}
\end{figure}

GPDs are non-diagonal matrix elements of a bilocal field
operator on the light cone. For partons with spin $1/2$ in a
pseudoscalar meson, the GPD associated with helicity conserving
partons is defined as follows \cite{diehl:review}: 
\begin{equation}
  \begin{split}
  &H^f(x,\xi,t)=\\
  &\frac{1}{2}\int\frac{\mathrm{d}z^-}{2\pi}e^{ix\tilde{P}^+z^-}\left.\langle\bar{P}^{\prime}
  |\bar{\psi}^f(-\frac{z}{2})\gamma^{+}\psi^f(\frac{z}{2})|\bar{P}\rangle\right|_{z^+=0,
  \mathbf{z_{\bot}}=\mathbf{0}}\,.
  \end{split}
  \label{eq:defgpd}
\end{equation} 
In this equation, $f$ refers to the flavor of the probed parton, while
$\tilde{P} = \frac{\bar{P}^{\prime}+\bar{P}}{2}$ and $t = \Delta^2 =
(\bar{P}^{\prime} - \bar{P})^2$. Definition
(\ref{eq:defgpd}) uses light-cone coordinates. A fourmomentum
$p=(p^0,p^1,p^2,p^3)$ can be written in light-cone coordinates as $p =
[p^+,p^-,\boldsymbol{p}_{\perp}]$ with $p^{\pm} = (p^0 \pm
p^3)/\sqrt{2}$ and $\boldsymbol{p}_{\perp} = (p^1,p^2)$. 
The skewedness $\xi$ and the average plus-momentum fraction of 
the struck parton $x$ are defined in fig. \ref{Fig:kinematics}:
$x$ denotes the fraction of the average meson plus-momentum that is
reabsorbed by the meson, while $\xi$ is a measure for the
plus-momentum that is lost in the process:  
\begin{equation}
  \xi = \frac{\bar{P}^+ - \bar{P}^{\prime+}}{\bar{P}^+ + \bar{P}^{\prime
      +}}\,.
\label{eq:defxi}
\end{equation}
These definitions coincide with the symmetrical variables that were
introduced by Ji \cite{ji:vars}. Their interpretation as plus-momentum
fractions holds in the infinite momentum frame. In literature, another
set of variables $(X,\zeta)$ has been introduced by Radyushkin
\cite{radyushkin:vars}, with $X=\frac{(x+\xi)}{(1+\xi)}$ and
$\zeta=\frac{2\xi}{(1+\xi)}$. Although Radyushkin's variables are more
closely related to the ones used in forward kinematics, we will use
Ji's choice of variables, as they reflect the symmetry between the
incoming and outgoing hadron states.

Definition (\ref{eq:defgpd}) is valid when the partons do not transfer
helicity. It only holds in a coordinate system where $\boldsymbol{q}$
and $\boldsymbol{\tilde{P}}$ are collinear and in the three-direction
\cite{ji:first}. From definition (\ref{eq:defgpd}), it is clear that
the GPDs are Lorentz invariant quantities. This means that, although
they are defined on the light cone, the GPDs can be calculated in any
convenient reference frame, provided that the average transverse
hadron momentum $\boldsymbol{\tilde{P}_{\perp}} =
\boldsymbol{0}$. Whenever a specific choice is needed, we will choose
a Breit frame which matches these conditions.

The skewedness $\xi$ is restricted to the interval $[0,\xi_{max}]$ with
\begin{equation}
  \xi_{max} = \sqrt{\frac{-t}{4 M^2 - t}}\,,
  \label{eq:ximax}
\end{equation}
where $M$ is the hadron mass. 

Stringent tests for any model calculation are found in the following
GPD properties. The first is the relation between the
electromagnetic form factor and the GPD,
\begin{equation}
  \int \mathrm{d} x H^f(x,\xi,t) = F^f(t)\,,
  \label{eq:ff_vs_gpd}
\end{equation}
where $\xi$ factors out and the partial electromagnetic form factors
$F^f$ are defined through 
\begin{equation}
F = \sum_f e_f F^{f}\,,
\label{eq:ff}
\end{equation}
with $e_f$ the electric charge of a parton with flavor $f$. The second
model constraint is called the polynomiality condition, which states
that the $n$th Mellin moment of the GPD is a polynomial in $\xi$ of
order $\le n$ \cite{ji:first}:
\begin{equation}
\int \mathrm{d}x~x^{n-1} H^f(x,\xi,t) =\sum_{i=0}^{n}a_i(t)\xi^i\,,
\label{eq:poly}
\end{equation}
where the coefficients $a_i$ of the polynomial depend on $t$. The
polynomiality condition can be regarded as a more general form of the
above form factor relation. Finally, for pions, isospin invariance and
charge conjugation lead to the following relation:  
\begin{equation}
  H_{\pi}^q(x,\xi,t)=-H_{\pi}^{\bar{q}}(-x,\xi,t)\,.
\label{Eq:isospinsymm}
\end{equation}
We will come back to these model constraints in sect. \ref{sec:results&disc}.

\section{\label{sec:bonnmodel}Formalism}
In the Bonn model, mesons are described as bound states of a
constituent quark and antiquark. The model, based on the
Bethe-Salpeter equation, is Poin\-car\'e covariant by construction.
Explicit covariance is important in the calculation of the GPD of
pseudoscalar mesons for two reasons. First, the pion is a
``deeply-bound'' boson whose static and dynamic properties are best
reproduced with a relativistic model \cite{metsch96}. Second, in the
calculation of dynamic quantities such as the GPD, recoil fourmomenta
have to be treated relativistically. The Bonn model is a few-parameter
model with only seven parameters which are fitted to the meson mass
spectrum \cite{koll00}. The fact that no relativistic corrections need
to be implemented is an enormous asset to minimize the amount of
parameters and maximize the predictive power. 
In refs. \cite{koll00,resag95,muenz94,muenzthesis}, the Bonn model is
described in detail. In sects. \ref{subsect:modelingr} and
\ref{subsect:reduction}, we give a brief summary to make this paper
more selfcontained.

\subsection{\label{subsect:modelingr}The model ingredients}
A meson with on-shell fourmomentum $\bar{P}$ is
described by the Bethe-Salpeter amplitude
\begin{equation}
\chi_{\bar{P}\alpha\beta}(x_1, x_2) = \langle\Theta|T\{\psi^{1}_{\alpha}(x_1)\bar{\psi}^{2}_{\beta}(x_2)\}|\bar{P}\rangle\,,\\
\label{eq:bs_ampl}
\end{equation}
where $T$ is the time ordering operator acting on the Heisenberg fermion
field operators $\psi_{\alpha}$. The indices in Dirac, flavor and
color space are combined in the multi-indices
$\alpha\equiv(\alpha,f,c)$. The variables $x_1$ and $x_2$ are the
fourvectors denoting the spacetime positions of the 
quark and the antiquark, respectively. 

\begin{figure}[t]
\centering
\includegraphics[width=0.8\columnwidth]{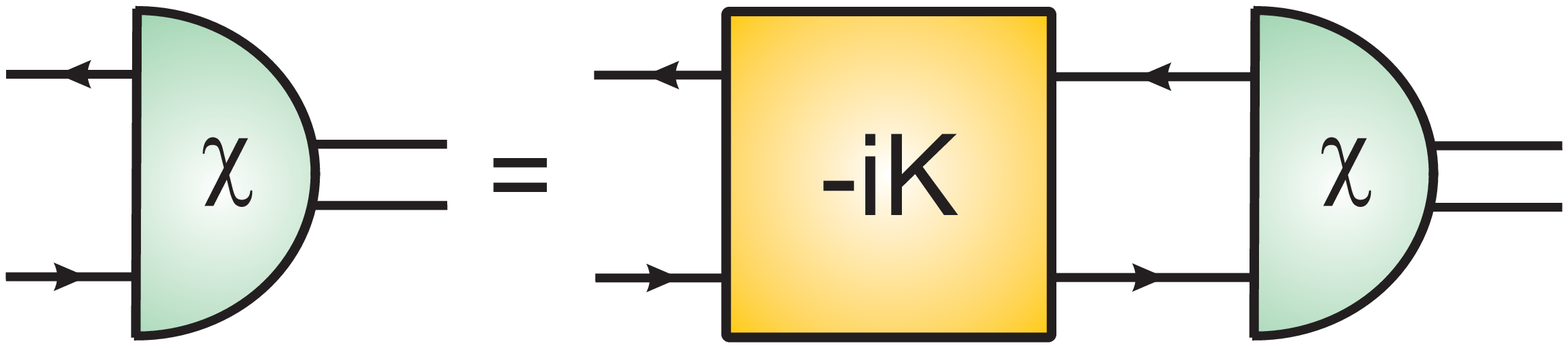}
\caption{\label{fig:bs_eq}Diagrammatic representation of the Bethe-Salpeter
  equation (\ref{eq:bs_eq}).}
\end{figure}

The Bethe-Salpeter amplitude (\ref{eq:bs_ampl}) is the solution of the Bethe-Salpeter
equation, which in momentum space reads \cite{resag95,bethesalpeter}:
\begin{equation}
  \chi_{\bar{P}} = -i G_{0\bar{P}}K_{\bar{P}}\chi_{\bar{P}}\,.
  \label{eq:bs_eq}
\end{equation}
Here, indices and arguments have been suppressed for notational 
simplicity. It is tacitly assumed that one integrates over
arguments and sums over indices that occur twice.
$G_{0\bar{P}}$ is the product of the one-particle propagators, while
$K_{\bar{P}}$ denotes the interaction kernel.
The pictorial representation of eq. (\ref{eq:bs_eq}) is shown in fig. \ref{fig:bs_eq}. 
The normalization of the Bethe-Salpeter amplitudes is given by
\cite{resag95}:
\begin{equation}
\bar{\chi}_{\bar{P}}\left[P^{\mu}\frac{\partial}{\partial P^{\mu}}(G_{0P}^{-1}
  + iK_P)\right]_{P=\bar{P}}\chi_{\bar{P}} = 2iM^2\,.
\label{eq:norm_cond}
\end{equation}
Note that eq. (\ref{eq:norm_cond}) is written in a frame independent way. 

To transform the Bethe-Salpeter equation (\ref{eq:bs_eq})
into a solvable integral equation, two \emph{Ans\"atze} are
made. The first assumption is the instantaneous approximation of the
interaction kernel $K_{\bar{P}}$,
\begin{equation}
K_{\bar{P}}(p,p^{\prime})\equiv V(p_{\perp\bar{P}},p_{\perp\bar{P}}^{\prime}),
\label{eq:inst_approx}
\end{equation}
where $p_{\perp P} \equiv p - (p \cdot P/P^2)P$ is the fourvector
perpendicular to $P$ with $p$ the relative fourmomentum between the
constituent quark and the constituent antiquark in the meson. Within
the instantaneous approximation, all retardation effects are
neglected. Strictly speaking, this is justified for a part of the
interaction (\textit{e.g.} the (colour-)Coulomb-interaction) in a
specific gauge (\textit{e.g.} the Coulomb gauge) only. Otherwise this
approximation will lead to a non-local and non-causal
model. Nevertheless, the model equations lead to frame-independent
results. Hence, the model can be called Poincar{\'e} covariant. The
second assumption is that the full quark propagators $S_j^F(p)$ can
suitably be approximated by free fermion propagators with an effective
constituent quark mass $m_j$,
\begin{equation}
S_j^F(p) \equiv \frac{i}{\feyndag{p} - m_j + i\epsilon}\,.
\end{equation}
The constituent quark masses are model parameters. Because isospin
symmetry is exact in the Bonn model, the calculation of the low-lying 
meson spectrum requires only two mass parameters: the non-strange and
the strange quark mass. 

\subsection{Reduction to the Salpeter equation\label{subsect:reduction}}
The Salpeter amplitude $\Phi(\boldsymbol{p})$ in the rest frame of
the bound state is defined as follows:
\begin{equation}
\Phi(\boldsymbol{p}) =
\int\frac{\mathrm{d}p^0}{2\pi}\chi_{\bar{P}}(p^0,\boldsymbol{p})\arrowvert_{\bar{P}=(M,\boldsymbol{0})}
\,.
\label{eq:defsalp}
\end{equation}
Making use of eqs. (\ref{eq:inst_approx}) and (\ref{eq:defsalp}), the
$p^0$-integration in eq. (\ref{eq:bs_eq}) can be carried out, leading
to the well-known Salpeter equation \cite{salpeter}: 
\begin{equation}
\begin{split}
\Phi(\boldsymbol{p}) = &\int\frac{\mathrm{d}^3p^{\prime}}{(2\pi)^3}
    \frac{\Lambda_1^-(\boldsymbol{p})\gamma^0[V(\boldsymbol{p},\boldsymbol{p^{\prime}})\Phi(\boldsymbol{p^{\prime}})]\gamma^0\Lambda_2^+(-\boldsymbol{p})}
    {M+\omega_1+\omega_2} \\
    - &\int\frac{\mathrm{d}^3p^{\prime}}{(2\pi)^3} \frac{\Lambda_1^+(\boldsymbol{p})\gamma^0[V(\boldsymbol{p},\boldsymbol{p^{\prime}})\Phi(\boldsymbol{p^{\prime}})]\gamma^0\Lambda_2^-(-\boldsymbol{p})}
    {M-\omega_1-\omega_2} \,,
\end{split}
\label{eq:salp_eq}
\end{equation}
with the energy projection operators $\Lambda_j^{\pm} = (\omega_j \pm
H_j)/(2\omega_j)$, the Dirac Hamiltonian $H_j(\boldsymbol{p}) =
\gamma^0(\boldsymbol{\gamma}\cdot\boldsymbol{p} + m_j)$ and the energy
$\omega_j = \sqrt{m_j^2 + |\boldsymbol{p}|^2}$ for the $j$th constituent
quark.

The potentials used in our calculations are the \emph{confinement
interaction} $\mathcal{V}$ and the \emph{'t Hooft instanton induced
interaction} $V_{III}$ \cite{tHooft}.  The confinement potential rises
linearly with the interquark distance and is multiplied by a Dirac
structure which reproduces the observed mass splittings in the meson
spectrum \cite{rickenthesis,anneliesthesis}:
\begin{multline}
\mathcal{V}(|\boldsymbol{x}_q - \boldsymbol{x}_{\bar{q}}|) =
\frac{1}{2} (a_c + b_c |\boldsymbol{x}_q - \boldsymbol{x}_{\bar{q}}|)
\\
\times (\Id \otimes \Id - \gamma^5 \otimes \gamma^5 - \gamma_\mu \otimes
\gamma^\mu) \; ,
\label{eq:conf_interaction}
\end{multline}
where $\boldsymbol{x}_{q(\bar{q})}$ is the position of the constituent
(anti-)quark. The instanton interaction
accounts for the mass splittings in the pseudoscalar and the scalar
sectors \cite{metsch96}. In momentum space, it can be written as
\begin{multline}
\int \frac{\mathrm{d}^3 p^{\prime}} {(2\pi)^3}
V_{III}(\boldsymbol{p},\boldsymbol{p}^{\prime})
\Phi(\boldsymbol{p}^{\prime}) = 4G(g,g^{\prime}) \\ \times \int
\frac{\mathrm{d}^3 p^{\prime}} {(2\pi)^3}
\mathcal{R}_{\Lambda}(\boldsymbol{p},\boldsymbol{p}^{\prime}) (\Id
Tr[\Phi(\boldsymbol{p}^{\prime})] + \gamma^5 Tr[\gamma^5
\Phi(\boldsymbol{p}^{\prime})]) \; ,
\label{eq:thooft_interaction}
\end{multline}
where $G(g,g^{\prime})$ includes the flavor dependent couplings $g$
and $g^{\prime}$, and $\mathcal{R}_{\Lambda}$ is a Gaussian regulating
function with cutoff $\Lambda$ \cite{rickenthesis,anneliesthesis}.
The 't~Hooft instanton induced interaction can account for the low
mass of the pion. The confinement and the instanton interaction
contain five additional model parameters, bringing the total to
seven. These seven model parameters are fitted to the Regge
trajectories and the pseudoscalar ground state masses and are kept
fixed in the calculation of dynamic observables. In this sense, the
results in sect. \ref{sec:results&disc} are predictions.

The Salpeter amplitudes $\Phi$ are calculated by
solving eq. (\ref{eq:salp_eq}). Imposing the normalization condition, written
in terms of the Salpeter amplitudes, then yields the mass spectrum \cite{resag95}.

\section{\label{sec:gpd_in_bm}GPDs in the Bethe-Salpeter Quark Model} 
\begin{figure}[t]
\centering
\includegraphics[width=0.8\columnwidth]{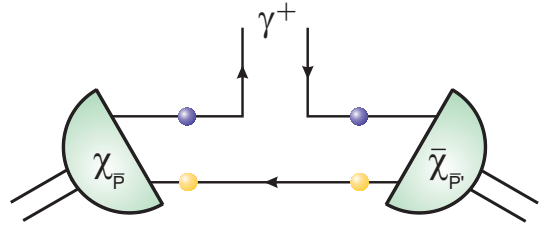}
\caption{\label{fig:coupling}Schematic representation of the
  second term in expression (\ref{eq:gpd_in_bm:matrixelement}) for the
  plus-component of the bilocal current matrix element.}
\end{figure}

The first step in the calculation of the generalized parton
distribution defined by eq. (\ref{eq:defgpd}) is the calculation of
the bilocal current matrix element
$\langle\bar{P}^{\prime}|\bar{\psi}(-\frac{z}{2})\gamma^{+}\psi(\frac{z}{2})|\bar{P}\rangle$.
In a Bethe-Salpeter based approach, the matrix element of any dynamic
variable can be calculated with the Mandelstam formalism
\cite{mandelstam}. This formalism acts as a starting point for the
derivation of the GPD in terms of the Bethe-Salpeter amplitudes. A
straightforward calculation yields the bilocal current matrix element
in lowest order \cite{anneliesthesis}:
\begin{equation}
\begin{split}
  &\langle\bar{P}^{\prime}|\bar{\psi}(-\frac{z}{2})\gamma^{+}\psi(\frac{z}{2})|\bar{P}\rangle\\
  = &\int\mathrm{d}^4x_2'~i\mathrm{Tr}\left\{\bar{\chi}_{\bar{P}^{\prime}}(x_2',\frac{z}{2})\left((i\feyndag{\partial}_{x_2'}-m_1)\chi_{\bar{P}}(x_2',-\frac{z}{2})\right)\gamma^{+}\right\}\\
  + &\int\mathrm{d}^4y_1'~i\mathrm{Tr}\left\{\left((i\feyndag{\partial}_{y_1'}-m_2)\bar{\chi}_{\bar{P}^{\prime}}(-\frac{z}{2},y_1')\right)\gamma^{+}\chi_{\bar{P}}(\frac{z}{2},y_1')\right\}\,.
\end{split}
\label{eq:gpd_in_bm:matrixelement}
\end{equation}
The second term in eq. (\ref{eq:gpd_in_bm:matrixelement}) (coupling to
the quark) is depicted in fig. \ref{fig:coupling}.
Inserting eq. (\ref{eq:gpd_in_bm:matrixelement}) in
eq. (\ref{eq:defgpd}) gives
\begin{equation}
  H(x,\xi,t) = H^{q}(x,\xi,t) +
  H^{\bar{q}}(x,\xi,t)\,,
  \label{eq:H}
\end{equation}
with the term originating from the quark-current coupling,
\begin{equation}
  \begin{split}
    &H^{q}(x,\xi,t) =\frac{i}{2}\int\frac{\mathrm{d}^4p}{(2\pi)^4}\delta\left(x\tilde{P}^+-\frac{\bar{P'}^+}{2}-p^+\right)\\
    &\times\mathrm{Tr}\left\{\left(-\frac{\feyndag{\bar{P}}}{2}+\feyndag{p}-m_2\right)\bar{\chi}_{\bar{P}^{\prime}}(p+\frac{\bar{P}^{\prime}}{2}-\frac{\bar{P}}{2})\gamma^{+}\chi_{\bar{P}}(p)\right\}\,,
  \end{split}
  \label{eq:Hq}
\end{equation}
and the term originating from the antiquark-current coupling,
\begin{equation}
  \begin{split}
    &H^{\bar{q}}(x,\xi,t)=\frac{i}{2}\int\frac{\mathrm{d}^4p}{(2\pi)^4}\delta\left(
    x\tilde{P}^+ + \frac{\bar{P'}^+}{2} -{p}^+ \right)\\
    &\times \mathrm{Tr}\left\{\bar{\chi}_{\bar{P}^{\prime}}(p+\frac{\bar{P}}{2}-\frac{\bar{P}^{\prime}}{2})\left(\frac{\feyndag{\bar{P}}}{2}+\feyndag{p}-m_1\right)\chi_{\bar{P}}(p)\gamma^+ \right\}\,.
  \end{split}
  \label{eq:Hqbar}
\end{equation}
Equations (\ref{eq:gpd_in_bm:matrixelement})-(\ref{eq:Hqbar}) make use
of the full Bethe-Salpeter amplitude $\chi$. This amplitude can be
reconstructed from the Salpeter amplitudes as follows: once
the Salpeter equation is solved, the vertex functions
$\Gamma_{\bar{P}} = G_{0\bar{P}}^{-1} \chi_{\bar{P}}$ can be
calculated. In the meson rest frame, these vertex functions read: 
\begin{equation}
\Gamma_{\bar{P}}(p_{\perp
  \bar{P}})\arrowvert_{\bar{P}=(M,\boldsymbol{0}) }= 
  \Gamma(\boldsymbol{p}) = -i \int\frac{\mathrm{d}^3p}{(2\pi)^3}[V(\boldsymbol{p},\boldsymbol{p^{\prime}})\Phi(\boldsymbol{p^{\prime}})]\,.
\label{eq:vert_func}
\end{equation}
By Lorentz boosting the resulting Bethe-Salpeter amplitude from the meson rest
frame to the frame in which the meson has on-shell momentum $\bar{P}$,
the amplitude in the latter frame is found: 
\begin{equation}
\chi_{\bar{P}}(p) = S_{\Lambda_{\bar{P}}}\chi_{(M,\boldsymbol{0})}(\Lambda_{\bar{P}}^{-1}p)S_{\Lambda_{\bar{P}}}^{-1}\,.
\label{eq:boost}
\end{equation}
In this equation, $\Lambda_{\bar{P}}$ denotes the Lorentz
transformation and $S_{\Lambda_{\bar{P}}}$ denotes the corresponding
boost operator acting on the fermion field operators in
eq. (\ref{eq:bs_eq}). Due to the instantaneous approximation, the
boost properties of the Bethe-Salpeter amplitude are described by the
above equation. All interaction dependence effectively enters the
boost operator through the bound state's mass.

Written in terms of these vertex functions, the quark GPD arising from
the Mandelstam formalism contains three quark propagators $S_F$:
\begin{equation}
\begin{split}
  H_{\pi}^{q}(x,\xi,t) = -\frac{1}{2}&\int\frac{\mathrm{d}^4p}{(2\pi)^4}\delta\left(\frac{2x+\xi-1}{2(1+\xi)}\bar{P}^+ -{p}^+ \right)\\
  \times~\mathrm{Tr}&\left\{\bar{\Gamma}_{\bar{P}^{\prime}}(p + \frac{\Delta}{2}) S_F^{1}(\frac{\bar{P'}}{2} + p + \frac{\Delta}{2})
  \gamma^+\right. \\
  &~~\left.S_F^{1}(\frac{\bar{P}}{2}+p)\Gamma_{\bar{P}}(p)S_F^{2}(-\frac{\bar{P}}{2} + p)\right\} \,,
\end{split}
\label{eq:Hq_vertex}
\end{equation}
These propagators can be directly linked with the intermediate quark
lines in the diagram of fig. \ref{fig:coupling}.  It turns out that
the denominators of these three propagator terms ensure the correct
support region for the GPDs, $x \in [-\xi, 1]$ (ERBL and DGLAP
regions) for $p^-$ independent vertex functions. For $p^-$ dependent
vertex functions, it is a priori unclear whether the GPD will be
confined to the support region.  In previous work, we have shown that
the Bonn model is prone to a support problem
\cite{vandyckvancauteren:letter}. In the next paragraphs, we will
quantitatively investigate which physical parameters influence the
support behavior of the model. The role of the variables $\xi$ and $t$
will be analyzed, and the dependence on the binding strength will be
examined through a comparison of the GPD of the kaon and the pion in
three different model variants.

\subsection{Model variants}
In the forthcoming section, the GPDs of pseudoscalar mesons will be
shown in three different model variants: the full model, the reduced
model and the increased quark mass (IQM) model. The parameters of
these models are presented in table \ref{ch:bonn_table:params}. Notice
that the parameters $a_c$ and $b_c$ of the confinement interaction
remain fixed for all three model variants.

\begin{table}
  \begin{center}
    \begin{tabular}{|c||c|c|c|}
      \hline
      Parameter & Full model  & Reduced model & IQM model\\
      \hline
      \hline
      $m_n$ [MeV] & ~380 & ~380& ~800 \\
      $m_s$ [MeV] & ~550 & ~550& ~-\\
      & & & \\
      $a_c$ [MeV] & -1135 & -1135& -1135\\
      $b_c$ [MeV/fm] & ~1300 & ~1300& ~1300\\
      & & & \\
      $g$ [GeV$^{-2}$] & 1.62 & 0.0& 1.62 \\
      $g^{\prime}$ [GeV$^{-2}$] & 1.35 & 0.0 & 1.35 \\
      $\Lambda$ [fm] & 0.42 & - & 0.42 \\
      \hline
      $M_\pi$ [MeV] & 141 & 572 & 1095 \\
      $\Delta_M^\pi$ [MeV] & 619 & 188 & 505 \\
      $\Delta_M^\pi/M_\pi$ & 4.39 & 0.33 & 0.46 \\
      \hline
      $M_K$ [MeV] & 506 & 728 & - \\
      $\Delta_M^K$ [MeV] & 424 & 202 & - \\
      $\Delta_M^K/M_K$ & 0.84 & 0.28 & - \\
      \hline
    \end{tabular}
  \end{center}
  \caption{Overview of the parameters of the three models that were
    used in this work: the constituent quark masses, the confinement
    offset and slope, the 't Hooft interaction range and the 't Hooft
    interaction strengths. Also presented is a summary of the masses
    $M$, binding energies $\Delta_M$ and relative binding energies
    $\Delta_M/M$ of the pion and the kaon in the different models.
  }
  \label{ch:bonn_table:params}
\end{table}

\subsubsection{Full model}
The full model is the one referred to as Model $\mathcal{B}$ in
ref. \cite{rickenthesis}. It provides an accurate description of the
pion and other meson properties such as its mass, electromagnetic form
factor, electroweak decay widths, etc. In the full model, the pion
mass is calculated as $M_{\pi}^{full} = 141$~MeV which implies a large
mass defect of $\Delta_{M_{\pi}}^{full} = (2m_n-M_{\pi}^{full}) =
619$~MeV; accordingly we shall call the pion ``deeply bound''. The
kaon is moderately bound in this model, with a mass of $M_{K}^{full} =
506$~MeV and a mass defect of
$\Delta_{M_{K}}^{full}=(m_n+m_s-M_{K}^{full}) = 424$~MeV.

\subsubsection{Reduced model}
In the reduced model, the 't Hooft instanton induced interaction is
omitted. As we have mentioned in sect. 3.2, the instanton interaction
accounts for the deep binding of the pion (and to a lesser degree also
of the kaon). Accordingly, neglecting the 't Hooft interaction will
provide insight into the importance of binding effects in the GPD
results. As a matter of fact, the calculated pion mass increases to
$M_{\pi}^{red} = 572$~MeV in this approach ($\Delta_{M_{\pi}}^{red} =
188$~MeV). The kaon mass increases to $M_{K}^{red} = 728$~MeV
($\Delta_{M_{K}}^{red} = 202$~MeV).

\subsubsection{Increased quark mass (IQM) model}
Not only the 't Hooft instanton induced interaction has an effect on
the pion binding energy. Also the non-strange constituent quark mass
$m_n$ affects the pion mass and mass defect. To investigate the
influence of the binding energy on the support of the generalized
parton distributions, both mechanisms must be studied.

In the IQM model, we will only show results for
the pion GPD. After combining the IQM model results with the kaon
results from the full model, one can determine the influence of the
heavy quarks. Increasing the non-strange quark mass by more than a
factor of $2$ to $m_n=800$~MeV
yields a pion mass of $M_{\pi}^{IQM}= 1095$~MeV. The mass defect in
this model is $\Delta_{M_{\pi}}^{IQM}=505$~MeV.

\subsubsection{Model summary}
The masses, binding energies and relative binding energies (defined as
the binding energy devided by the mass) calculated in the different
models are summarized in table \ref{ch:bonn_table:params}. The deep
binding of the pion in the full model is reflected in the high
relative binding energy. $\Delta_M/M$ in the full model is much
smaller for the kaon than for the pion. Further, $\Delta_M/M$ of the
kaon in the full model is larger than in the reduced model, and also
larger than $\Delta_M/M$ of the pion in the reduced and the IQM
models.  In sect. 5, we will come back to these (relative) binding
energies.

\subsection{Model constraints}
In sect. 2, we introduced three constraints that serve as
stringent tests for any GPD calculation. These are the isospin
symmetry relation (\ref{Eq:isospinsymm}) for the pion GPD, the form
factor relation (\ref{eq:ff_vs_gpd}) and the polynomiality condition
(\ref{eq:poly}). In this section, we elaborate on these constraints,
and show that they are fulfilled in our model.

The pion up and down quark GPDs must fulfill the isospin symmetry
relation of eq. (\ref{Eq:isospinsymm}). The equality is exact in our
calculations. The strange quark content of the kaon prevents an
isospin symmetry relation of the type (\ref{Eq:isospinsymm}).
written.  The results show a small difference between the quark and
antiquark GPDs at opposite $x$ (\emph{e.g.} in the full model,
$H_{K^+}^u(x=0.5,\xi=0,t=-0.5$~GeV$^2)$ $= 0.764$ versus
$H_{K^+}^s(x=-0.5,\xi=0,t=-0.5$~GeV$^2)$ $= -0.784$). We will
elaborate on these differences in sect. \ref{sec:results&disc}.

The GPD is related to the electromagnetic form factor through relation 
(\ref{eq:ff_vs_gpd}). Taking into account eq. (\ref{eq:ff}), which
relates the partial form factors with the meson form factor, and the
isospin symmetry relation (\ref{Eq:isospinsymm}) for the pion, one finds
that 
\begin{equation}
\int_{-\infty}^{+\infty} \mathrm{d}x H_{\pi^+}^u(x,\xi,t) = F_{\pi^+}(t) \,.
  \label{ch:results_eq:ff_vs_gpd_pion}
\end{equation}
Note that, due to the support properties of the GPDs in the Bonn
model, the integration domain is $x\in(-\infty,+\infty)$. For the
kaon, the relation becomes  
\begin{equation}
\begin{split}
e_u \int_{-\infty}^{+\infty} \mathrm{d}x &H_{K^+}^u(x,\xi,t) \\
&+  e_s
 \int_{-\infty}^{+\infty} \mathrm{d}x H_{K^+}^s(x,\xi,t) = F_{K^+}(t) \,, 
  \label{ch:results_eq:ff_vs_gpd_kaon}
\end{split}
\end{equation}
where $e_u=2/3$ and $e_s=-1/3$.
 
We have compared the results of
eqs. (\ref{ch:results_eq:ff_vs_gpd_pion}) and
(\ref{ch:results_eq:ff_vs_gpd_kaon}) with a direct computation of the
electromagnetic form factors $F_{\pi^+}$ and $F_{K^+}$ in the Bonn
model \cite{koll00}. These numerical calculations were performed
independently and yielded results which were compatible at the few-\%
level \cite{anneliesthesis}. When discussing these form factors
results, it is worth stressing that the instantaneous approximation
does not result in current non-conservation. Although the EM current
in lowest order in the instantaneous approximation generally does not
fulfill the Ward-Takahashi identity, it does so in the case of a $0^-
\to 0^-$ elastic matrix element~\cite{muenzthesis}.

An even more stringent test than the form-factor condition of
eq. (\ref{eq:ff_vs_gpd}), is the polynomiality condition of
eq. (\ref{eq:poly}). This condition was verified numerically for
different values of $t$ up to order $n=5$ \cite{anneliesthesis}.

In~\cite{noguera06}, the authors point out that the parton
distribution (PD) of the pion should fulfill four conditions: (a) the
support region is $-1 < x < 1$, (b) the first moment of the GPD must
be 1 (normalization) for any value of $/xi$, (c) in the case of the
pion the second moment of the PD must be $1/2$ (momentum conservation)
and (d) isospin and momentum conservation imply that the PD of the
pion must be symmetric under the change $x \to (1-x)$. As to condition
(a), the pion vertex function in our model is expressed in Minkowski
spacetime and has an intricate analytic structure as opposed to
Ref.~\cite{noguera06}. It was already argued in our previous
work~\cite{vandyckvancauteren:letter} that this can lead to a support
problem. Condition (b) is theoretically fulfilled in our model and has
been checked numerically~\cite{anneliesthesis}. Condition (c) is not
fulfilled in our model and may be attributed to large off-shell
effects. Concerning condition (d), isospin symmetry translates into
condition~(\ref{Eq:isospinsymm}) in our model, and this has been
confirmed numerically to be valid for every value of $\xi$ and $t$.

\section{\label{sec:results&disc}Results and Discussion}
\begin{figure}[h!]
\centering
\includegraphics[width=0.5\textwidth]{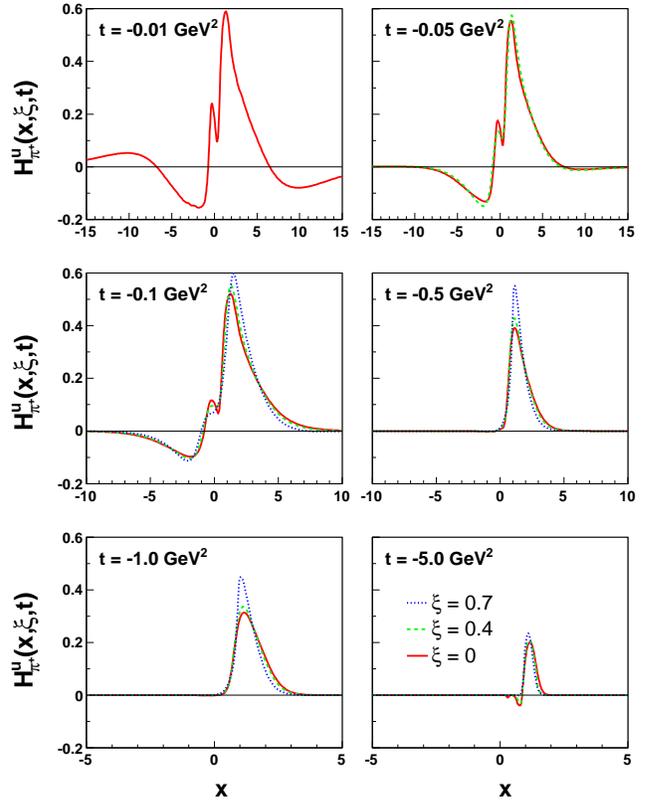}
\caption{\label{ch:results_fig:pion_fullmodel} The pion
  GPD $H_{\pi^+}^u$ in the full model for different values of $t$. Values  
  for $\xi$ shown are $\xi = 0$ (red, solid line), $\xi=0.4$ (green,
  dashed line) and $\xi=0.7$ (blue, dotted line).} 
\end{figure}

\begin{figure}[h!]
\centering
\includegraphics[width=0.5\textwidth]{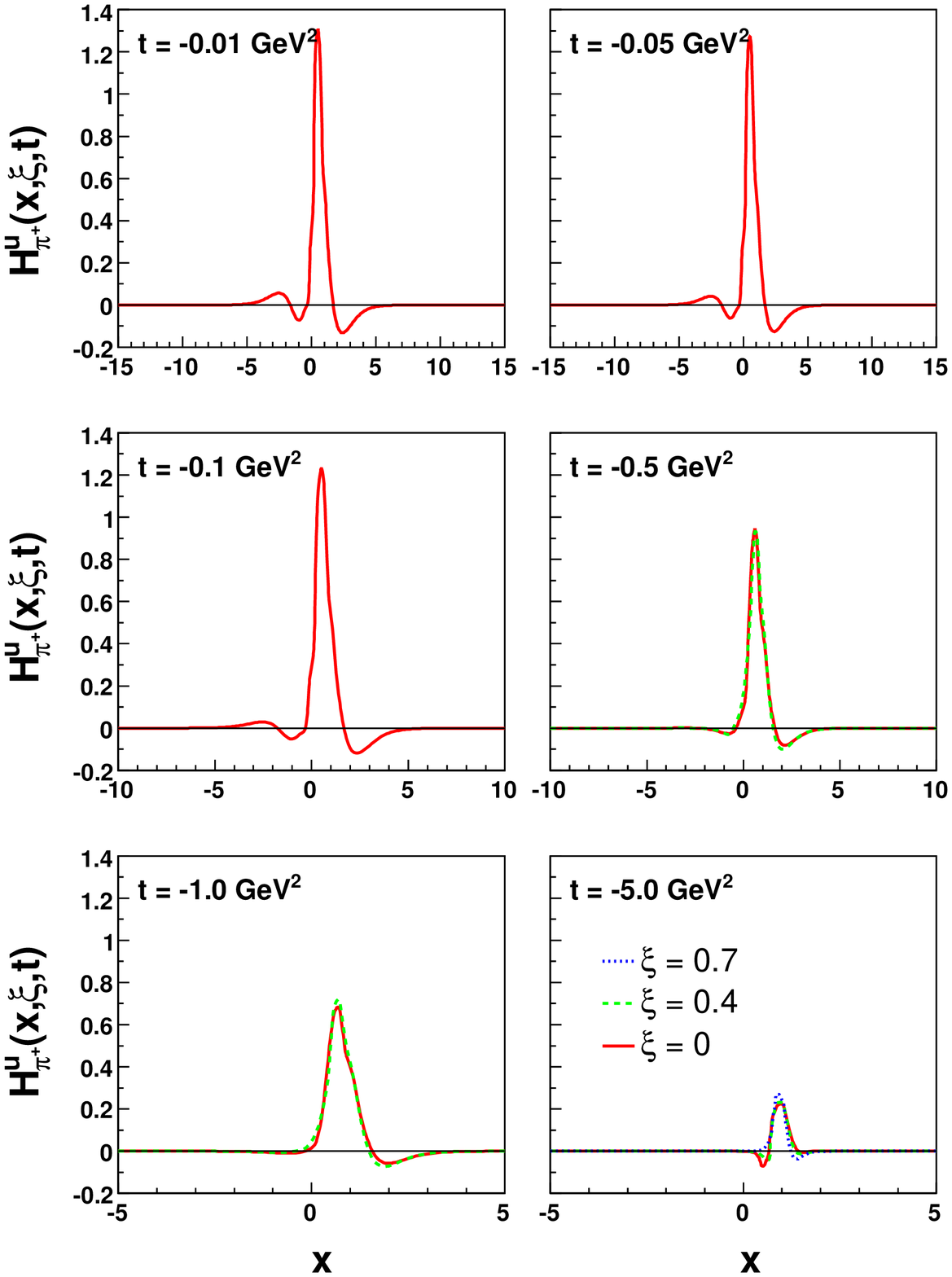}
\caption{\label{ch:results_fig:pion_reducedmodel} The pion
  GPD $H_{\pi^+}^u$ in the reduced model for different values of $t$. Values  
  for $\xi$ shown are $\xi = 0$ (red, solid line), $\xi=0.4$ (green,
  dashed line) and $\xi=0.7$ (blue, dotted line).} 
\end{figure}

\begin{figure}[h!]
\centering
\includegraphics[width=0.5\textwidth]{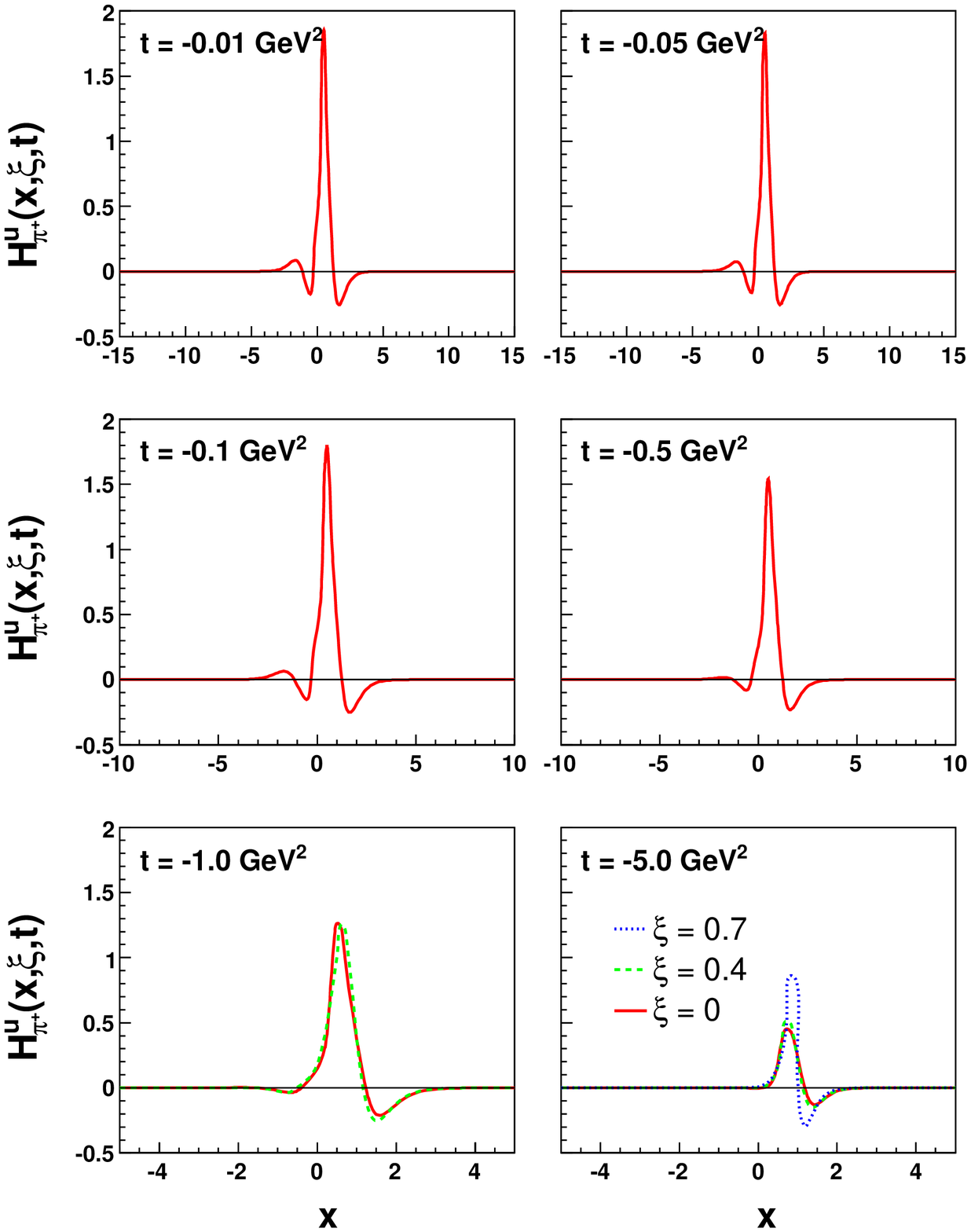}
\caption{\label{ch:results_fig:pion_highermnmodel} The pion
  GPD $H_{\pi^+}^u$ in the IQM model for different values of $t$. Values  
  for $\xi$ shown are $\xi = 0$ (red, solid line), $\xi=0.4$ (green,
  dashed line) and $\xi=0.7$ (blue, dotted line).} 
\end{figure}

\begin{figure}[h!]
\centering
\includegraphics[width=0.5\textwidth]{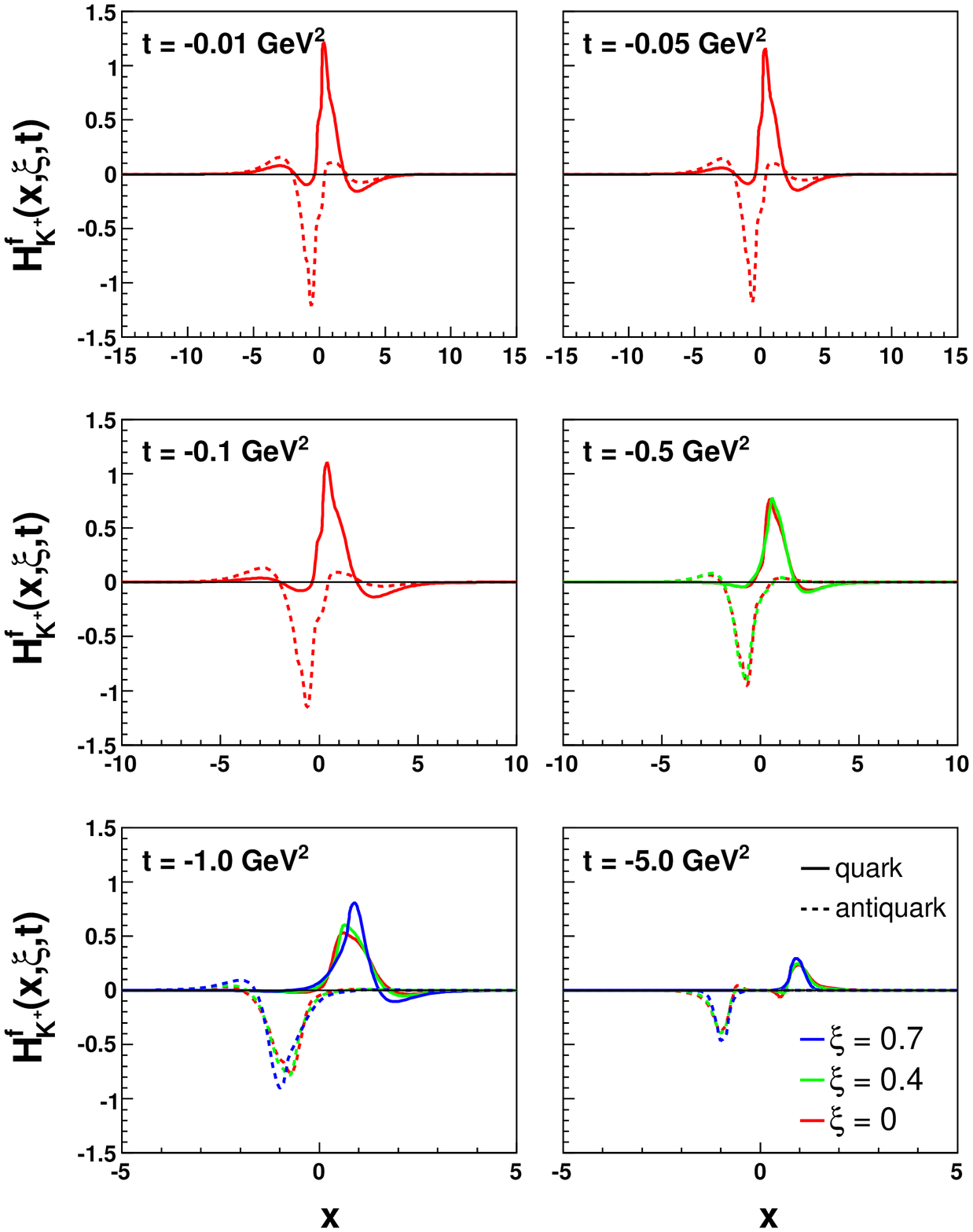}
\caption{\label{ch:results_fig:kaon_fullmodel} The kaon
  GPDs $H_{K^+}^u$ and $H_{K^+}^s$ in the full model for different values
  of $t$. Values for $\xi$ shown are $\xi = 0$ (red line), $\xi=0.4$
  (green line) and $\xi=0.7$ (blue line). The solid line refers to the
  $u$ GPD, the dashed line to the $\bar{s}$ GPD.} 
\end{figure}

\begin{figure}[h!]
\centering
\includegraphics[width=0.5\textwidth]{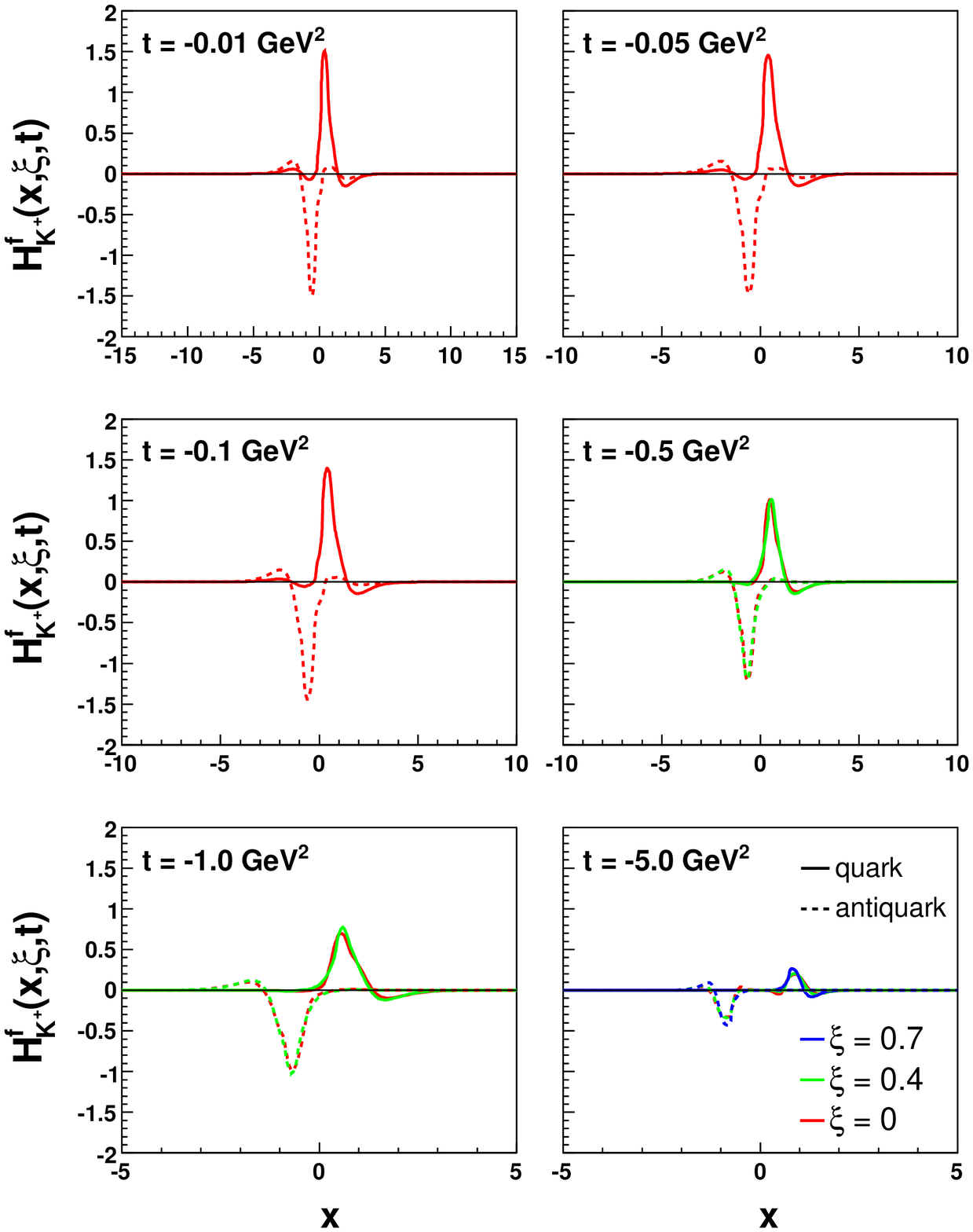}
\caption{\label{ch:results_fig:kaon_reducedmodel} The kaon
  GPDs $H_{K^+}^u$ and $H_{K^+}^s$ in the reduced model for different values
  of $t$. Values for $\xi$ shown are $\xi = 0$ (red line), $\xi=0.4$
  (green line) and $\xi=0.7$ (blue line). The solid line refers to the
  $u$ GPD, the dashed line to the $\bar{s}$ GPD.}
\end{figure}

\begin{figure}[t]
\centering
\includegraphics[width=0.5\textwidth]{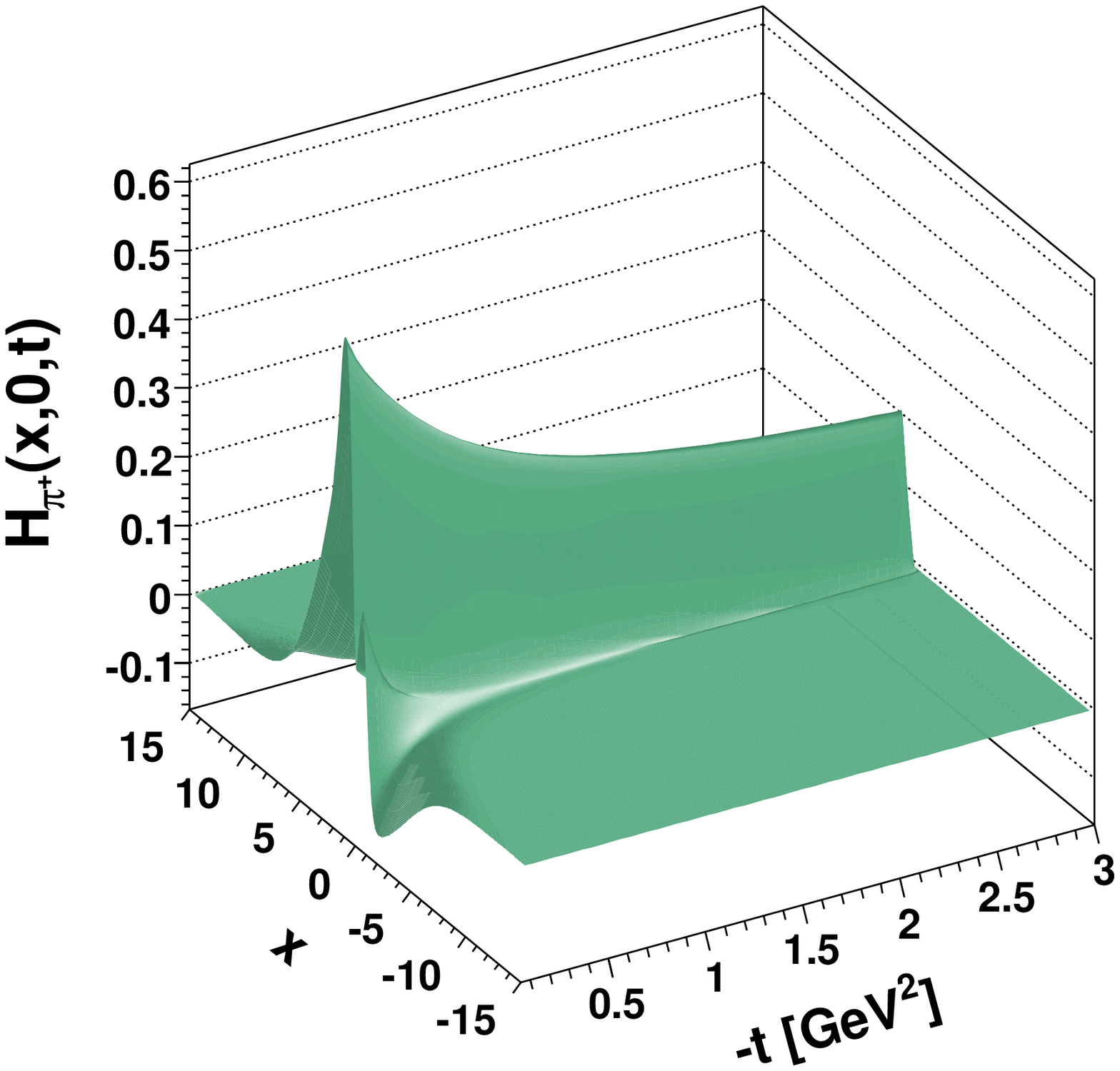}
\caption{The $H_{\pi^+}^u(x,0,t)$ GPD in the full model as a function
  of $x$ and $-t$ for $\xi=0$.
  \label{ch:results_fig:3d}}
\end{figure}

The $\pi^+$ up-quark GPD results in the three different model variants
are shown in figs. \ref{ch:results_fig:pion_fullmodel} -
\ref{ch:results_fig:pion_highermnmodel}, the $K^+$ up-quark and
strange-antiquark GPDs in figs. \ref{ch:results_fig:kaon_fullmodel} -
\ref{ch:results_fig:kaon_reducedmodel}. A three-dimensional picture of
the (full model) $\pi^+$ up-quark GPD as a function of $x$ and $t$
with skewedness $\xi=0$ is presented in
fig. \ref{ch:results_fig:3d}. For all cases, the results are shown for
a representative selection of $t$ and $\xi$ values.

It was shown in ref. \cite{vandyckvancauteren:letter} that GPDs in the
Bonn Model violate the support condition. This can also be deducted
from figs. \ref{ch:results_fig:pion_fullmodel} -
\ref{ch:results_fig:kaon_reducedmodel}.
In general, the curves tend to have longer tails as $|t|$
decreases. Especially for the deeply bound pion in the full model,
this effect is clearly visible (see fig. \ref{ch:results_fig:3d}). As
the relative binding energy decreases, the effect becomes
smaller. Furthermore, whereas the maximum of the GPD curve lies
outside the support region for the deeply bound pion in the full model
(peak value at $|x|>1$), it lies within the support interval for the
other pion and kaon calculations.

Another observation which can be made, is that
the pion GPD in the full model displays a clear shoulder in the region
$x \in [0,1]$ at $|t| \lesssim 0.1$~GeV$^2$ (see
fig.~\ref{ch:results_fig:pion_fullmodel}). This shoulder is much less
pronounced or even absent in the other results. In this respect, it is
interesting to note that the computed pion form factor shows a bump at
low $Q^2 = -t$, which can be ascribed to the instantaneous
approximation~\cite{koll00,kollthesis,muenz95}. The observed shoulder
in the pion GPD might therefore be an artefact of this
\textit{Ansatz}.

A direct comparison with other model calculations
(\cite{noguera:bsapp,tibmil1}) is not straigthforward due to the
different behavior with respect to the support properties. A
double-peaking behavior for $x=(1 \pm \xi)/2$ in the reduced or IQM
model variants is not seen. In both of these models, however, the
binding energy represents a significant portion of the meson mass:
about $1/3$ ($1/2$) for the reduced (IQM) model (see
Table~\ref{ch:bonn_table:params}). This is in contrast with the
calculations of Ref.~\cite{noguera:bsapp}, where the meson mass is
very close to the sum of the quark masses and the double-peaking
behavior emerges.

The observed model dependence of the GPD properties points at a
significant relative binding energy dependence of the support 
of the generalized parton distributions in the Bonn model. To quantify
the support problem, a support parameter ($\phi$) is introduced. For
the pion, the generalized quark and antiquark distributions are
related via the isospin symmetry relation
(\ref{Eq:isospinsymm}), so that the knowledge of one of them
implies the knowledge of the other. The support parameter is therefore
defined through the quark GPD:
\begin{equation}
\phi = \frac{\int_{-\xi}^1
  |H_{\pi^+}^{u}(x,\xi,t)|\mathrm{d}x}{\int_{-\infty}^{\infty}
  |H_{\pi^+}^{u}(x,\xi,t)|\mathrm{d}x}\,. 
\label{ch:results_eq:def_supportparameter}
\end{equation}
$\phi$ is a measure for the relative support violation: when the
correct kinematical regions are resolved, $\phi = 1$. The smaller
$\phi$, the worse the support. For the kaon, the isospin symmetry
relation (\ref{Eq:isospinsymm}) is not valid and different support
parameters are introduced for the quark and antiquark GPDs:
\begin{equation}
\phi^q = \frac{\int_{-\xi}^1
  |H_{K^+}^{u}(x,\xi,t)|\mathrm{d}x}{\int_{-\infty}^{\infty}
  |H_{K^+}^{u}(x,\xi,t)|\mathrm{d}x}
\end{equation}
and
\begin{equation}
\phi^{\bar{q}} = \frac{\int_{-1}^{\xi}
  |H_{K^+}^{s}(x,\xi,t)|\mathrm{d}x}{\int_{-\infty}^{\infty}
  |H_{K^+}^{s}(x,\xi,t)|\mathrm{d}x}\,. 
\label{ch:results_eq:def_supportparameter_kaon}
\end{equation}
The difference between $\phi^q$ and $\phi^{\bar{q}}$ will be used to
investigate the flavor dependence of the support problem.

\label{sect:results_sub:supportparameter}
\begin{table*}[t]
  \begin{center}
    \begin{tabular}{|c c||c|c|c|c|c|c|}
      \hline
      & & \multicolumn{6}{c|}{$t$ (GeV$^2$)}\\
      Model & $\xi$ & $-0.01$ & $-0.05$ & $-0.1$ & $-0.5$ &
      $-1.0$ & $-5.0$ \\ 
      \hline
      \hline
      Full model & 0 & 0.08 & 0.11 & 0.13 & 0.21 & 0.20 & 0.17\\
      & 0.4 & - & 0.12 & 0.14 & 0.21 & 0.22 & 0.17\\
      & 0.7 & - & - & 0.12 & 0.17 & 0.20 & 0.12 \\
      Reduced model & 0 & 0.60 & 0.62 & 0.64 & 0.69 & 0.69 & 0.63 \\
      & 0.4 & - & - & - & 0.70 & 0.71 & 0.61 \\
      & 0.7 & - & - & - & - & - & 0.56 \\     
      IQM model & 0 & 0.66 & 0.67 & 0.67 & 0.71 & 0.73 & 0.68 \\
      & 0.4 & - & - & - & - & 0.75 & 0.70 \\
      & 0.7 & - & - & - & - & - & 0.64 \\
      \hline
    \end{tabular}
  \end{center}
  \caption{Values of the pion support parameter $\phi$ of
    eq. (\ref{ch:results_eq:def_supportparameter}) corresponding 
    to figs. \ref{ch:results_fig:pion_fullmodel} -
    \ref{ch:results_fig:pion_highermnmodel}. \label{ch:results_table:full}}  
\end{table*}    
Table \ref{ch:results_table:full} lists the values of the support
parameter $\phi$
for the pion GPD results of figs.~\ref{ch:results_fig:pion_fullmodel} -
\ref{ch:results_fig:pion_highermnmodel}. Table
\ref{ch:results_table:full_kaon_qbar} shows the values of the
parameters $\phi^q$ and $\phi^{\bar{q}}$
belonging to the kaon curves of
figs. \ref{ch:results_fig:kaon_fullmodel} and
\ref{ch:results_fig:kaon_reducedmodel}.

\begin{table*}[t]
  \begin{center}
    \begin{tabular}{|c c c||c|c|c|c|c|c|}
      \hline
      & & & \multicolumn{6}{c|}{$t$ (GeV$^2$)}\\
      Parameter & Model & $\xi$ & $-0.01$ & $-0.05$ & $-0.1$ & $-0.5$ &
      $-1.0$ & $-5.0$ \\ 
       \hline
      \hline
      $\phi^q$ & Full model & 0 & 0.46 & 0.48 & 0.50 & 0.58 & 0.61 & 0.47 \\
      & & 0.4 & - & - & - & 0.60 & 0.62 & 0.48 \\
      & & 0.7 & - & - & - & - & 0.56 & 0.56 \\
      & Reduced model & 0 & $0.68$ & $0.69$ & $0.70$ & $0.73$ & $0.73$ & $0.64$ \\
      & & 0.4 & - & - & - & 0.76 & 0.75 & 0.62 \\
      & & 0.7 & - & - & - & - & - & 0.63 \\
      $\phi^{\bar{q}}$ & Full model & 0 & 0.43 & 0.45 & 0.47 & 0.56 & 0.61 & 0.42 \\
      & & 0.4 & - & - & - & 0.58 & 0.61 & 0.43 \\
      & & 0.7 & - & - & - & - & 0.55 & 0.57 \\
      & Reduced model & 0 & 0.65 & 0.66 & 0.67 & 0.72 & 0.73 & 0.65 \\
      & & 0.4 & - & - & - & 0.73 & 0.74 & 0.64 \\
      & & 0.7 & - & - & - & - & - & 0.69 \\      
      \hline
    \end{tabular}
  \end{center}
  \caption{Values of the kaon quark and antiquark support parameters
    of eq. (\ref{ch:results_eq:def_supportparameter_kaon})
    corresponding to figs. \ref{ch:results_fig:kaon_fullmodel} -
    \ref{ch:results_fig:kaon_reducedmodel}. \label{ch:results_table:full_kaon_qbar}}   
\end{table*}    

For the pion GPD, it is seen from table \ref{ch:results_table:full}
that the support is much better in the reduced and the IQM models than
in the full model. In going from the full model to either of the two
other models in which the pion is less deeply bound, the support
parameter increases significantly (\emph{e.g.} from $0.08$ in the full
model to $0.60$ in the reduced and $0.66$ in the IQM model for
$t=-0.01$~GeV$^2$ and $\xi=0$). We observed in the previous paragraph
that, especially for the pion in the full model, the GPD curves tend
to broaden with decreasing $|t|$. Table \ref{ch:results_table:full}
shows that this behavior can be attributed to the support properties
of the curves. For fixed skewedness $\xi$, the support improves for
increasing $|t|$ except for very large values ($|t| \gtrsim
5.0$~GeV$^2$). For a fixed $|t|$, the $\phi$ exhibits hardly any
dependence on $\xi$.

Also for the kaon GPD, the support improves when the instanton induced
interaction is switched off. Notice that the values of the two support
parameters $\phi^q$ and $\phi^{\bar{q}}$ are similar. This result is
compatible with the finding that the quark and antiquark GPD have
similar shapes and shows that the small differences between both GPDs
do not alter the support significantly.  Combining these observations
with the fact that the support does improve significantly in the IQM
model with respect to the full model, we conclude that concerning the
support the binding strength is more important than the particular
constituent quark mass or flavor.

\section{\label{sec:Sum&concl}Summary and Conclusions}
In this work, the isoscalar twist-two generalized parton distributions
of the pion and the kaon were calculated in the Poincar\'e covariant
Bethe-Salpeter constituent quark model developed by the Bonn
group. The first moment of the GPD in the full $x$-region equals the
electromagnetic form factor. Results were shown in the full $x$-region
for several values of $\xi$ and $t$. It turns out that the Bonn model
violates the support condition. We have illustrated the strong
correlation between the support and the (relative) binding energy of
the meson.  Therefore, the deep binding of the pion induces strong
support violations. We have found that $\xi$ hardly influences the
support, whereas a moderate dependence on $t$ is predicted. The
constituent quark masses have an impact on the support, but only
through the (relative) binding energy. A mass difference between
constituent quarks hard\-ly affects the corresponding GPD curves.

The reason for the violation of the support condition in the Bonn
constituent quark model is not fully understood: It might be related
to the instantaneous approximation made. Although this assumption is
formally covariant, the neglect of retardation effects could lead to
wrong analytic properties of the hadronic vertex functions, in
particular for ``deeply-bound'' states such as the pion. Some hints in
this direction were already found earlier in the calculated pion
electromagnetic form factor, which showed a conspicuous unphysical
structure at small momentum transfers~\cite{kollthesis,muenz95}. The
present detailed analysis provides additional support for this
conjecture.

The observed support properties of the Bonn constituent quark model
could also be interpreted as follows. As the Bonn model is a
\emph{soft scale} model, with neither on-shell particles nor
asymptotic freedom properties, the support violation can be considered
as a manifestation of soft scale physics. This prohibits the use of
the available QCD-based evolution equations to the model. In contrast,
a soft-scale model with the correct support properties can be
evolved. It should be clear, however, that also in this case, the use
of the evolution equations relies on an extrapolation of perturbative
QCD to soft scales. Additionally, it should be noted that the Mellin
moments of the GPDs at $\xi = 0$ can be evolved, even in the case of a
support violation.

\subsection*{Acknowledgements}
  The authors wish to thank F. LLanes Estrada, S. Scopetta and D. Van
  Neck for enlightening discussions. AVD and TVC are grateful to the
  Research Foundation - Flanders (FWO) for financial support. BCM
  acknowledges the support of the European Community-Research
  Infrastructure activity under the FP6 ``Structuring the European
  Research Area'' programme (Hadron Physics, contract number
  RII3-CT-2004-506078) and the support within the DFG SFB/TR16
  ``Subnuclear Structure of Matter - Elektromagnetische Anregung
  subnuklearer Systeme''.

\bibliographystyle{epj}
\bibliography{references.bib}
\end{document}